\documentclass[conference,a4paper,10pt]{IEEEtran}
\IEEEoverridecommandlockouts

\usepackage{cite}
\usepackage{url,amsmath,amsfonts,amssymb,mathrsfs,epsfig,bm}
\usepackage{graphicx,comment}
\usepackage{color}
\def\R{{\mathbb R}}

\def\E{{\mathbb E}}
\def\P{{\mathbb P}}
\def\ut{\underline{t}}

\def\Scal{\mathcal{S}}

\def\Pcal{\mathcal{P}}

\def\eg{{\em e.g.},~}
\def\ie{{\em i.e.},~}
\def\cf{{\em cf.}~}
\newtheorem{proposition}{Proposition}[section]

\newcommand{\beqa}{\begin{eqnarray*}}
\newcommand{\eeqa}{\end{eqnarray*}}
\newcommand{\be}{\begin{eqnarray}}
\newcommand{\ee}{\end{eqnarray}}

\usepackage{caption}
\usepackage{subcaption}
\usepackage{fullpage}
\usepackage{times}
\usepackage{xcolor}
\usepackage{fancyhdr,amsmath,amssymb}
\usepackage[ruled,vlined,nofillcomment]{algorithm2e}
\usepackage{mathrsfs}
\usepackage{booktabs}

\SetCommentSty{mycommfont}
\include{pythonlisting}

\newcommand{\qed}{\raisebox{.65ex}{\fbox{\rule{0mm}{0mm}}} ~\\}

\def\BibTeX{{\rm B\kern-.05em{\sc i\kern-.025em b}\kern-.08em
    T\kern-.1667em\lower.7ex\hbox{E}\kern-.125emX}}

\begin{document}

\title{On a Caching System with Object Sharing\thanks{This
research was supported in part by NSF CNS 
grants 1526133 and 1717571 and by a Cisco
Systems URP gift.}}

\author{
\begin{tabular}{cc}
G. Kesidis, N. Alfares, X. Li, B. Urgaonkar, M. Kandemir & T. Konstantopoulos\\
School of EECS & Dept of Mathematics\\
Penn State University & University of Liverpool\\
University Park, PA, USA, 16802 & Liverpool, UK, L69 7ZL\\
\{gik2,nna5040,xzl45,buu1,mtk2\}@psu.edu & 
t.konstantopoulos@liverpool.ac.uk
\end{tabular}
}


\maketitle

\begin{abstract}
We consider a content-caching system that
is shared by a number of proxies. The cache could
be located in an edge-cloud datacenter and the
proxies could each serve a large population
of mobile end-users. Each proxy operates  its
own LRU-list of a certain capacity in the shared cache.
The length of objects simultaneously 
appearing in plural LRU-lists is
equally divided  among them, \ie object sharing among
the LRUs. 
We provide a ``working-set" approximation for this system
to quickly estimate the cache-hit probabilities under
such object sharing, which can be used to facilitate 
admission control.  Also, a way to 
reduce ripple evictions, \ie {\tt set} request overhead,
is suggested.
We give numerical results for our 
MemCacheD with Object Sharing (MCD-OS) prototype.
\end{abstract}

\section{Introduction}

As the public cloud-computing marketplace rapidly expands and diversifies
its services and associated pricing rules,
the edge-cloud marketplace is still developing. 
Comparable edge-cloud services most likely will be much more expensive
than those of the public (``remote") cloud.
In particular, content-caching services preferably
locate at the edge to reduce networking costs and delays.
One way to reduce their costs is to have
different caching proxies (as, \eg \cite{squid})
share objects stored in common cache memory.

We herein 
consider $J$ proxies
that each service a large pool of users/processes
making requests for content 
from a database with $N$ data-objects via
a cache of size $B\ll N$ memory units,  \eg caching as 
part of a Content Distribution Network (CDN). 
Each proxy typically operates under a Least Recently
Used (LRU) caching policy wherein the most recently queried
for data-objects are cached.
The $J$ proxies {\em share} both cache memory and
possibly also the upload network-bandwidth to their users.
Note that for caching of encrypted data (\eg owing
to copyright protections), a layered encryption
strategy (as in block chains, legers)
could be used to first encrypt to the
network edge and then encrypt to the individual
(authorized) users.

In this paper, we consider a caching system where
the noncooperative proxies each pay for an allocation of
cache memory (and possibly network I/O as well), thus preventing
starvation of any proxy.
Objects may be shared among different LRU-lists (or just ``LRUs",
each corresponding
to a proxy)  as \cite{Stoica16}.
That is, the cost of storing a common object in the LRUs is shared
among the proxies.
Also, an
LRU-list miss but physical cache hit 
is accompanied by a  delay corresponding to a physical cache miss.
This said, a proxy may make inferences regarding the LRU-lists
of others by comparing the cache hits they experience to what they
would be without object sharing.
Mock queries may change some near-future LRU-list  misses to hits 
(particularly for content not in the physical cache), but will come at the cost of 
both memory and network I/O resources (possibly {\em causing}
some near-future cache misses that would have been hits).
So, the free-riding  behavior
described in \cite{Stoica16} is disincentivized. 

\noindent{\bf Summary Contributions:}
We give an accurate working-set approximation for the 
cache-hit  probabilities of the caching-with-object-sharing
system of \cite{Stoica16}. 
We implement this system in Memcached in part to practically ascertain
the object sharing's cache-eviction ``ripple effect."
We also study the ripple-eviction problem and describe an approach
to overbooking the caches that mitigates it.

This paper is organized as follows.
Related prior work is discussed in Section \ref{sec:related}.
In Section \ref{sec:spatial}, an approach to cache memory management
is presented wherein a cached object's length is shared among multiple
LRU-lists. 
In Section \ref{sec:shared-cache-analysis}, we propose an approach
to approximating hitting times for such a system of shared
cache memory under the Independent Reference Model (IRM) model.
Numerical results are given in Section \ref{sec:numer}.
In Section \ref{sec:mcd-os}, we described an implementation of
MemCacheD with Object Sharing (MCD-OS) and give additional
numerical results.
Finally, we conclude with a summary  and discussion of future
work in Section \ref{sec:concl}.

\section{Related prior work}\label{sec:related}

There  is substantial prior work on cache sharing,
including at the network edge in support of 
mobile end-users, \eg
\cite{Golrezaei12,YWang16,Tassiulas19}.
At one extreme, the queries of the proxies are aggregated
and one LRU cache is maintained for all of them using the
entire cache memory. At another extreme, the cache memory
is statically partitioned among the proxies (without object
sharing). 
For example, 
\cite{Towsley17} describes how cache memory can be partitioned
according to a game wherein different proxy utilities increase
with cache-hit probability. 
In our object-sharing problem formulation (involving non-cooperative
users of a for-profit caching service), cache memory is not
statically partitioned, but  there is ``virtual" 
cache memory allocated per user each of which is used for a LRU-list of
potentially shared data objects.

For a system with a single LRU (LRU-list) in the cache,
a lower priority (paying less)  proxy
could have a different
tail (least recently used object) pointer corresponding to lower
amount of allocated memory, but different proxies would then
compete for ``hot" (higher ranked) objects stored in the cache.
To reduce such competition,
an interesting system of \cite{Eyilmaz19}
also has a single LRU maintained in the cache
but with highest priority (paying most) proxies having access to 
the entire cache
while lower priority proxies having a {\em head} (most recently
used  object)
pointer corresponding
to a lower amount of allocated memory. 

Now consider a scenario where the clients of 
different proxies may query for (\eg via a {\tt get} request in Memcached) the same object.
In \cite{Stoica16},  proxies are assigned a share of
cached content based on their demand. Individual data
objects are {\em shared} among different proxy caches 
that store them, each according to the LRU policy,
\ie a share of their length is attributed to each proxy's 
cache (LRU-list). 
In \cite{Stoica16},
the cache blocks some requests selected at random to deter
a proxy from ``cheating" by issuing mock requests for 
specific content primarily of interest only to its
users in order to keep it cached (hot),
while leveraging cached content apportioned to other proxies,
\ie more generally popular content,
recall the discussion of Section 1.

\section{Object-sharing in cache memory}\label{sec:spatial}

Suppose cache memory is ``virtually" allocated
 so that proxy $i\in\{1,2,...,J\}$ {\em effectively} receives $b_i\leq B$
amount of memory.  
Each partition is managed simply  by a LRU linked-list of pointers 
(``LRU-list" or just ``LRU" in the following) to
objects stored in (physical) cache memory collectively for all
the proxies.

Let $\Pcal(n)\subset [J]$ be the set
of proxies for which object $n$ currently appears in their LRU-list, 
where
$\Pcal(n)=\emptyset$ if and only if
 object $n$ is not {\em physically} cached.
Note that $\Pcal(n)$ is not disclosed to the proxies, \ie the
proxies cannot with certainty tell whether objects {\em not} in
their LRU-list are in the cache. 

Upon request by proxy $i$ for object $n$ of length $\ell_n$,
object $n$ will be placed at the head of $i$'s LRU-list
and all other objects in LRU-list $i$ are 
demoted in rank.

If the request for object $n$ was a hit on LRU-list $i$, then
nothing further is done.

If it was a miss on LRU-list $i$, then 
\begin{itemize}
\item if the object is not stored in the physical
cache then it is fetched from the database, stored in the
cache and forwarded to proxy $i$; 
\item otherwise, the object is produced for proxy $i$ after
an equivalent delay.
\end{itemize}
Furthermore, add $i$ to $\Pcal(n)$
(as in  \cite{Stoica16}), \ie
\be
\Pcal(n) & \leftarrow & \Pcal(n)\cup\{i\}, \label{increase-C}
\ee
then  add the length $\ell_n/|\Pcal(n)|$   to
LRU-list $i$ and reduce
the ``share" of all other caches containing $n$ 
to $\ell_n/|\Pcal(n)|$  
(from $\ell_n/(|\Pcal(n)|-1$)).  

So, if the query for (get request of) object $n$ by proxy $i$ is a miss,
its LRU-list length will be inflated and
possibly exceed its allocation $b_i$;
thus, LRU-list eviction of its tail (least recently
used) object may be required.
When an object $m$ is  ``LRU-list evicted" by any proxy,
the apportionment of $\ell_m$ to other LRU-lists
is {\em increased} (inflated), which may cause other objects to
be LRU-list evicted by other proxies.
A simple mechanism that the cache operator could use is to
evict until no LRU-list exceeds its allocated memory is
to iteratively:
\begin{enumerate}
\item identify the LRU-list $i$ with largest overflow (length minus allocation)
\item if this largest overflow is not positive then stop
\item evict $i$'s lowest-rank  object
\item reassess the lengths of all caches
\item go to 1.
\end{enumerate}
This is guaranteed to terminate after a finite number of iterations
because in every iteration,
one object is evicted from an LRU-list and there are obviously only
ever a finite number of objects per LRU-list.

\begin{figure*}
     \centering
     \begin{subfigure}[t]{0.475\textwidth}
         \centering
         \includegraphics[width=\textwidth]{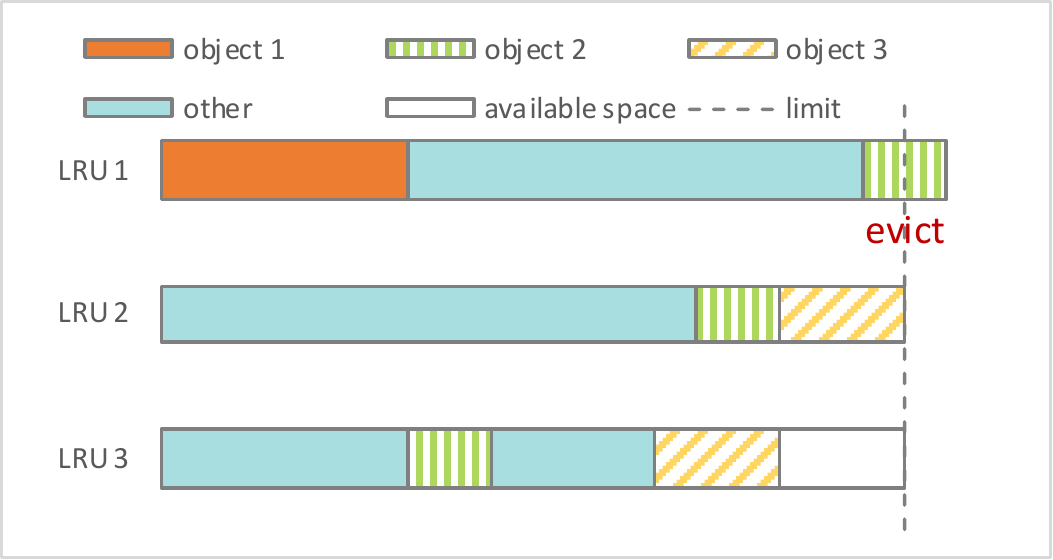}
         \caption{Assume we have three equal sized LRUs. Object 2 is shared by LRU 1, 2, and 3. Object 3 is shared by LRU 2 and 3. A new item, object 1, is inserted to the head of LRU 1.}
         \label{fig:inflation-1}
     \end{subfigure}
     \hfill
     \begin{subfigure}[t]{0.475\textwidth}
         \centering
         \includegraphics[width=\textwidth]{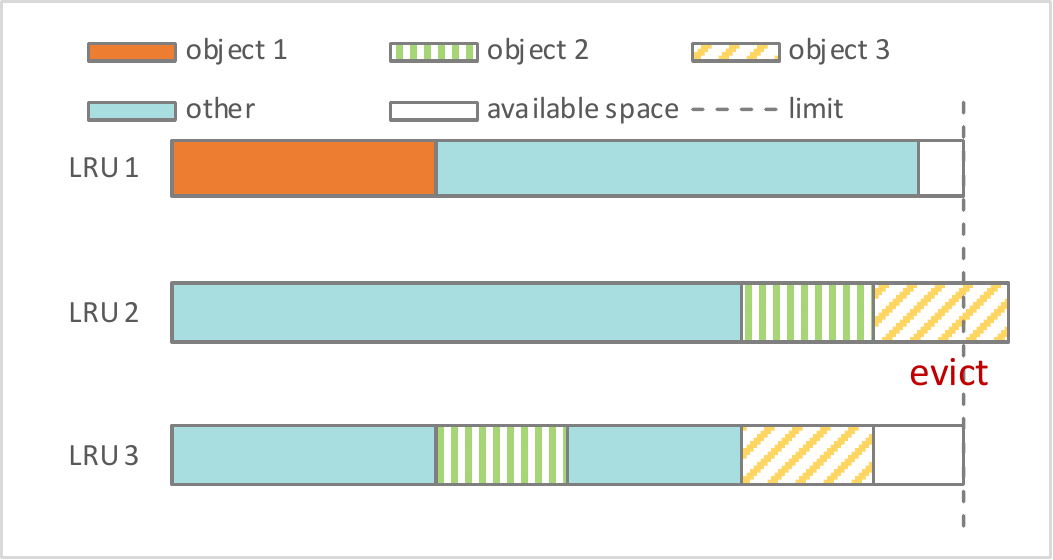}
         \caption{LRU 1 evicts object 2. So, the virtual length of object 2 
inflates in LRUs 2 and 3. So, LRU 2 exceeds its limit and needs to evict
object 3.}
         \label{fig:inflation-2}
     \end{subfigure}
     \\
     \begin{subfigure}[t]{0.475\textwidth}
         \centering
         \includegraphics[width=\textwidth]{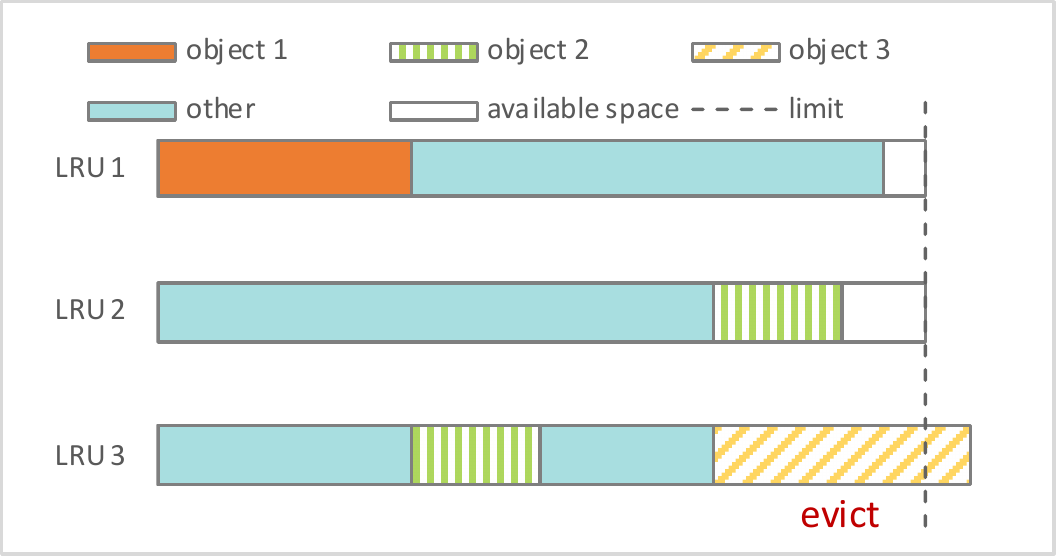}
         \caption{The increased virtual length of object 3 similarly requires
LRU 3 to evict.}
         \label{fig:inflation-3}
     \end{subfigure}
     \hfill
     \begin{subfigure}[t]{0.475\textwidth}
         \centering
         \includegraphics[width=\textwidth]{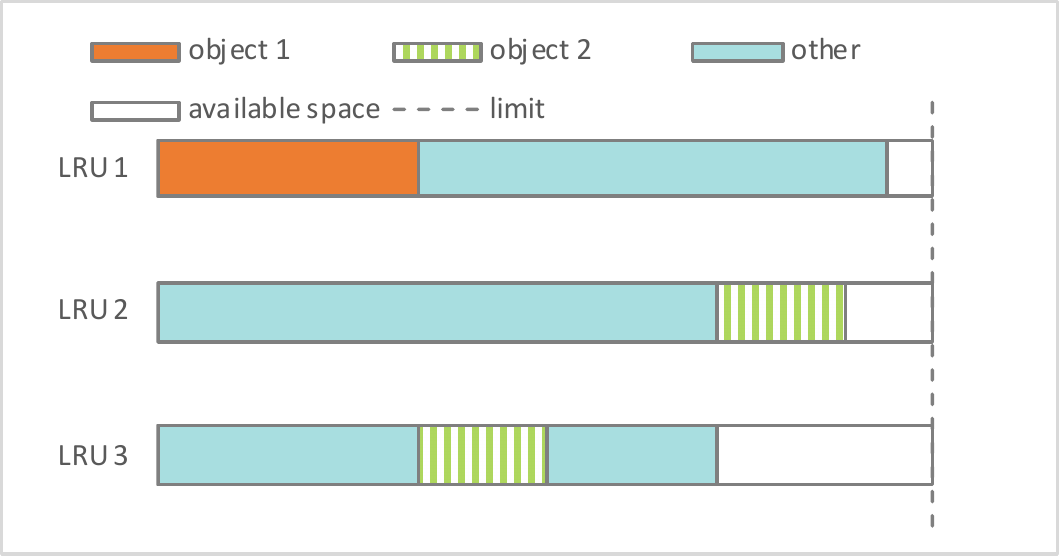}
         \caption{LRU 3 evicts object 3. Now no LRU exceeds its limit and 
processing of the insertion of object 1 into LRU 1 in (a) is completed.}
         \label{fig:inflation-4}
     \end{subfigure}
        \caption{Inflation of object size caused by deleting a shared object.}
        \label{fig:inflation}
\end{figure*}

Figure~\ref{fig:inflation} provides an illustration of such inflation for a shared
cache serving 3 LRUs. 
Here, insertion of a new object causes evictions in all three LRUs.

As another example,
consider a scenario where object $x$ is in LRU-list $j$ but not $i$ and
object $y$ is in both but at the tail of $i$. Also, both caches are
full.  So, a query for $x$ by $i$
(LRU miss but cache hit) causes $i$  to evict $y$.
Thus, from $j$'s point-of-view, $x$ deflates but $y$ inflates,
so evictions from $j$ may or may not be required.

Also, a ``set" request for an object simply {\em updates} an object 
in the cache which may cause it to inflate and, in turn, cause evictions.
Though we do not consider set requests in this paper, our implementation
does accommodate them, \cf Section \ref{sec:mcd-os}.

Note that if  during the eviction iterations,
$\Pcal(n)\rightarrow \emptyset$ for some object $n$, 
then $n$ may be removed from the physical cache (physically evicted) --
cached objects $n$ in the physical cache that are not
in any LRU-lists are flagged as such and have lowest priority
(are first evicted if there is not sufficient room for any object
that is/becomes a member of any LRU-list).
Even under LRU-list eviction consensus, the physical cache
may store an object {\em if it has room} to try to avoid having 
to fetch it again from the database in the future. 


In summary, a single proxy $i$ can cause a new object $n$ to enter the cache
($\Pcal(n)$ changes from $\emptyset$ to $\{i\}$) whose entire length $\ell_n$ is
applied to its cache memory allocation $b_i$,
but a consensus is required for an object $n$ to leave the cache 
($\Pcal(n)\rightarrow \emptyset$).
So, the physical cache itself is not LRU.
Also, as objects are requested, their apportionments to proxy LRU-lists
may deflate and inflate over time.

\begin{proposition}\label{prop:better-than-not-sharing}
For a fixed set of  active proxies $i$,
this object-sharing caching system  will have
a higher stationary object hit-rate per proxy compared to a 
{\em not-shared} system of LRU caches, where each proxy $i$'s LRU cache
has the same amount of allocated memory  ($b_i$) in both cases.
\end{proposition}

An elementary proof of this proposition
is based on  a simple coupling argument 
to show that for each proxy, the objects in  the
not-shared system's cache
are always a subset of what's in the LRU-list of the shared system.
This follows simply because the size of any object $n$ 
apportioned to the shared system
$$\ell_n/|\Pcal(n)|\leq \ell_n,$$
\ie not greater than
 its full size which is apportioned in the system without
object sharing.

\section{Approximating LRU-list hit probabilities under the IRM}\label{sec:shared-cache-analysis}

\subsection{Working-set approximations for shared-object caches under IRM}

In this section, we propose an approach to computing the 
approximate hitting probabilities of the foregoing caching system
following the
Denning-Schwartz ``working-set approximation"
\cite{DS72} for a not-shared cache under the IRM.
\cite{Fagin77,Fricker12} nicely address the
asymptotic accuracy of this approximation.
Also see \cite{Che02}. 

Let $\lambda_{i,k}$ be the mean request rate for 
object $k$, of length $\ell_k$, by proxy $i$.
A simple generalization of the working-set
approximation for variable-length objects is:
if $\min_i b_i \gg \max_k \ell_k$ then 
\be
\label{ws-approx}
\forall i~ ~
b_i = \sum_{k=1}^N h_{i,k} \ell_k 
\ee
where  
\be
\forall i,k~~ h_{i,k}  = 1-\mbox{e}^{-\lambda_{i,k} t_i}
\label{h_i-approx}
\ee
and $t_i$ are interpreted as (assumed common)
mean eviction times of objects $k$ in LRU-list $i$, 
\ie the time between when an object enters the cache and
when it's evicted from the cache.

For our shared caching system, only a fraction of an object
$k$'s length $\ell_k$ will be attributed to a particular
LRU-list $i$, depending on how $k$  is shared over (eviction) time $t_i$.
For all $i,k$, let  this attribution be 
$L_{i,k} \leq \ell_k$, \ie
\be
\forall i, ~b_i & = & \sum_{k=1}^N h_{i,k}L_{i,k} 
=\sum_{k=1}^N (1-\mbox{e}^{-\lambda_{i,k} t_i})L_{i,k}.
\label{ws-approx-shared}
\ee
One may take
\be\label{L-def1}
L_{i,k}^{(1)} & = & \ell_k \E \frac{1}{1+\sum_{j\not=i} Z_{j,k}} ,
\ee
where $Z_{j,k}$ are {\em independent} Bernoulli random variables  such
that $h_{j,k} = \P(Z_{j,k}=1)=1-\P(Z_{j,k}=0)$.
That is, under the assumption of independent LRU-lists,
$L_{i,k}^{(1)}$ is the stationary mean attribution of 
the length of object $k$ 
to LRU-list $i$  given that $k$ is stored in LRU-list $i$.
For example, 
for a system with just $J=2$ caches, 
\ie $j\in\{1,2\}$, 
\begin{align*}
\E \frac{1}{1+\sum_{j\not=i} Z_{j,k}} 
& =  1\cdot (1-h_{3-j,k}) +\frac{1}{2} h_{3-j,k} \\ 
& =   1-\frac{1}{2}h_{3-j,k}.
\end{align*}
So, substituting (\ref{L-def1}) into
(\ref{ws-approx-shared}) gives, for $i\in\{1,2\}$,
\beqa
 0 =   b_i
 - \sum_{k=1}^N (1-\mbox{e}^{-\lambda_{i,k} t_i})
( 1-\frac{1}{2}(1- \mbox{e}^{-\lambda_{3-i,k} t_{3-i}}))
\ell_k  ;
\eeqa
a system with two nonlinear equations in two unknowns $t_1,t_2$.

Empirically, we found that using (\ref{L-def1}) is a good estimate
of when $J>2$, \cf Section \ref{sec:numer}, but significantly under-estimates
the object hitting probabilities, \ie $L_{i,k}^{(1)}$ is too large,
when $J=2$. 
To explain this, we argue that object sharing creates a kind
of {\em positive
association} between  the LRU-list hit events, because hits in one 
cause the objects to effectively reduce in size in others,
so that they remain in the LRU-lists longer (larger eviction times), thus
increasing the hit probabilities in others.

To see why, consider the simple Prop. \ref{prop:F}  below
for Boolean random variables $Y_{j,k}$ indicating  the {\em dependent}
events that object  $k$ is stored in LRU-list $j$ in steady-state.


\begin{proposition}\label{prop:F}
For an arbitrary object index $k$,
consider $J\geq 2$  nonnegative random variables $Y_{1,k},\ldots,Y_{J,k}$ 
and $J$ 
other 
random variables
$Z_{1,k},\ldots,Z_{J,k}$ such that, for all $i$,  $Z_{i,k}$ and $Y_{i,k}$ have the same
distribution.
If for any LRU-list $i\in\{1,2,\ldots,J\}$ we have
\be
\label{POD}
\P\left( \sum_{j\not=i} Y_{j,k}  \le x\right) \le \P\left(\sum_{j\not=i}Z_{j,k} \le x\right)
\ee
then
\be\label{gyeqgz}
\E \left(1+\sum_{j\not=i} Y_{j,k}\right)^{-1} & \leq & \E \left(1+\sum_{j\not= i} Z_{j,k}\right)^{-1}.
\ee
\end{proposition}
\noindent {\bf Proof:}
Let $F_1, F_2$ respectively be the CDFs of 
$ \sum_{j\not=i} Y_{j,k}$.
So by hypothesis, $F_1(x) \le F_2(x)$ for all $x \in \R$.
Let $F_i^{-1}(u) := \inf\{x \in \R:\, F_i(x) > u\}$, $i=1,2$.
By change of variables in Lebesgue-Stieltjes integrals, we have
\[
\int_{-\infty}^{\infty}
 g(x)\mbox{d}F_i(x) 
=
\int_0^1 g(F_i^{-1}(u)) \mbox{d}u.
\]
Since $F_1 \le F_2$ we have $F_1^{-1} \ge F_2^{-1}$
and so $g(F_1^{-1}(u))  \le g( F_2^{-1}(u))$ for all $0 < u <1$.

Finally, take $g(x)=1/(1+x)$.
\qed

Note that, for our purposes herein, the random variables
$Z_{j,k}$ are independent.

Also note that according to (\ref{POD}),
$\sum_{j\not=i} Y_j$ tends to be larger than
$\sum_{j\not=i} Z_j$, similar to
positive associations
or positive correlations properties among random variables $Y_i\geq 0$
\cite{KS81,J-DP83,Wajc17}.

Substituting (\ref{L-def1}) 
into
(\ref{ws-approx-shared}) gives, for $i\in\{1,2,\ldots,J\}$,
{\small
\be
 0 =  b_i - \sum_{k=1}^N h_{i,k}
 \E \frac{1}{1+\sum_{j\not=i} Z_{j,k}} 
\ell_k  
~ =:  \frac{\partial u_i}{\partial t_i} =: \partial_i u_i 
\label{t_i-equations}
\ee
}
Under (\ref{h_i-approx}) and 
$\E Z_{j,k}=h_{j,k}$ for
independent Boolean $Z_{j,k}$, equations
(\ref{t_i-equations}) are a set of $J$ equations in
$J$ unknowns $\{t_i\}_{i=1}^J$.

Note that 
for all the above definitions,
$\forall i,k, ~L_{i,k}\leq \ell_k$, so one
expects corresponding hit cache probabilities 
to be larger than without object-sharing; recall
Prop. \ref{prop:better-than-not-sharing}.

\subsection{Existence and uniqueness of solution to
the working-set approximation (\ref{t_i-equations})}

A basic assumption is that,
\be \label{b_i-bound}
\forall i~b_i  <  \frac{1}{J}\sum_{k=1}^N \ell_k,
\ee
\ie no LRU-list  is large enough to hold all of the
objects even if the objects were fully shared.

\begin{proposition}\label{prop:approx}
If (\ref{b_i-bound}) holds then
there are real numbers  $s_j\geq 0, S_j<\infty$, such that
$s_j<S_j$ and 
there exists a unique
solution $\{t_i\}_{i=1}^J\in\prod_{i=1}^J[s_i,S_i]$ 
to (\ref{t_i-equations}).
\end{proposition}

\noindent {\bf Proof:}
Consider the quantities $u_i$ as utilities of a noncooperative 
$J$-player game
with strategies 
$$\{t_j\}_{j=1}^J \in \prod_{j=1}^J[s_j,S_j]=:\Scal$$ 
where $0\leq s_j<S_j<\infty$.
First note that each $u_i$ of (\ref{t_i-equations})
is continuously differentiable on $\Scal$. 

For a $J$-dimensional vector $\ut=(t_1,\ldots,t_J) \in \Scal$ 
let $\ut_{-i}$ be the $(J-1)$-vector obtained by eliminating the entry $t_i$.
Since the strategy-space $\Scal$ is compact 
and the utility functions $u_i(\ut)$ are strictly concave in $t_i$
(since $\partial_i^2 u_i < 0$) a Nash equilibrium exists
\cite{basar99dynamic}.
Alternatively, we can use Brouwer's theorem \cite{Border85}
to establish existence of the Nash equilibrium.

Generally, a Nash equilibrium may occur on the boundary of the strategy-space.
However, note here that for an arbitrary 
$\ut_{-i}$, 
{\small
\begin{align*}
\lim_{t_i \rightarrow 0} \partial_i u_i(t_i ,\ut_{-i}) & =  b_i >  0
~~\mbox{and}\\
\lim_{t_i \rightarrow \infty} \partial_i u_i(t_i ,\ut_{-i}) & = 
b_i - 
\sum_{k=1}^N \frac{1}
{1+\sum_{j\not= i} (1-\mbox{e}^{-\lambda_{j,k} t_j})} \ell_k \\
& \le b_i - \frac{1}{J}\sum_{k=1}^N \ell_k 
< 0, \qquad\mbox{by (\ref{b_i-bound})}.
\end{align*}
}
Because of this and the strict concavity of $u_i$ in $t_i$,
if all $S_j$ are sufficiently large and $s_j\geq 0$ sufficiently small,
then all $\partial_i u_i (\ut)$ are {\em unimodal} in $t_i$ for
all $\ut_{-i}$ such that $\ut \in \Scal$. As a result, the Nash equilibria
are all interior to $\Scal$ so that the
first-order necessary conditions for $u_i$-optimality must all hold, \ie
(\ref{t_i-equations}) are satisfied.

By such unimodality and  because
strict concavity implies
$\partial_i^2 u_i  \not = 0$, 
uniqueness of  the solution  follows.
\qed



Note that the diagonal-dominance conditions implying 
negative definiteness of the Jacobian of the gradient map,
which would imply uniqueness of the solution 
$\{t_i\}_{i=1}^J$ to (\ref{t_i-equations})
\cite{Rosen65,Moulin84},
do not hold here.

\subsection{Discussion: overbooking with shared objects}\label{sec:static}

Consider a caching system as described above with LRU-lists
but {\em without} object sharing, \ie the full length of 
an object is charged to each LRU-list in which it resides. 
In this case, from the proxies' point-of-view, the system is
just as static cache partitioning as mentioned in Section \ref{sec:related}.
Suppose LRU/proxy $i$ is paying to experience 
the cache-hit probabilities it would
get if cache memory amount $b_i^*$ was dedicated to it without object sharing,
\ie $b_i^*$ is prescribed in the Service Level Agreement (SLA).
Consider a {\em virtual} cache memory allocation 
$b_i$ given by (\ref{ws-approx-shared}) and (\ref{L-def1})
(accurate for $J>3$ LRUs, \cf Section \ref{sec:numer}).
Specifically, let $h_{i,n}$ be the cache hitting probability
of object $n$ under object sharing (so depends on $b_i$ - recall
 (\ref{ws-approx}))
and $h_{i,n}^*$ be that without object sharing
(so depends on $b_i^*$); and 
define minimal $b_i$ such  that
\be\label{h-bounds}
\forall i,n,~  h_{i,n}\geq h_{i,n}^* & \Rightarrow &
 \forall i, ~ b_i \leq b_i^*.
\ee

Object sharing with $J$ LRUs operates so that 
\be\label{virtual_allocations}
\sum_{i=1}^J b_i & \leq & B ,
\ee
which allows for the possibility  of {\em overbooking}, \ie
\be\label{overbooking}
\sum_{i=1}^J b_i^* & > & B .
\ee

For purposes of admission control, before the degree of
object-sharing of a new LRU $J+1$ can be assessed, 
the cache operator can  conservatively admit a new proxy  $J+1$ if
\begin{eqnarray}
b_{J+1}^* &\leq & B- \sum_{i=1}^J b_i,
\label{overbooking-ac}
\end{eqnarray}
where the cloud operator estimates the ``virtual" cache
allocation $b_i$ for existing proxies  $i\in\{1,2,...,J\}$.
Once admitted, the object popularities $\lambda_{i,n}$ can be estimated
and fed into our working-set approximation to compute cache-hit probabilities
under object-sharing
toward  determining the proper virtual allocation  $b_i$\footnote{Note that
virtual allocations may also need to be recomputed when LRUs ``depart" 
the cache.}.
Alternatively, 
the object cache-hit probabilities can be directly estimated
by simply trial reducing virtual allocation $b_{J+1}$ starting from $b_{J+1}^*$.
Or, LRU $J+1$ can be less conservatively
admitted based on a virtual allocation correponding to some estimated
object popularities based on those of existing LRUs $1,...,J$.

\subsection{Discussion: Reducing Ripple Evictions (RRE)}\label{sec:os-rre}

Obviously, if two LRUs $i\not = j$ have very similar demand, then they will tend to have many of the same objects cached, so 
will tend to have larger ripple-eviction effects,
recall Figure \ref{fig:inflation} and \cf Section \ref{sec:ripple-overhead}.
If the same objects in different LRUs are being evicted at 
approximately the same time, 
then a policy of delayed {\em batch} evictions may be effective at RRE.

Consider an allocation $\hat{b}_i$ to proxy $i$ satisfying  
$b_i \leq \hat{b}_i < b_i^*.$ Another approach  is to
give back $\sum_i \hat{b}_i-b_i$ in memory savings
in order to reduce ripple evictions.
Let a {\em primary} eviction on an LRU be one that
is  caused when a cache miss occurs on it.
Primary evictions occur when LRU $i$ exceeds $b_i$,
but {\em ripple} evictions occur only when LRU $i$ exceeds $\hat{b}_i$.
So, LRU $i$ may at times exceed $b_i$ (but never exceeds $\hat{b}_i$).
Note that under object sharing, some LRUs $i$ may shrink in size when a miss occurs for LRU $j\not=i$ involving an object in LRU $i$ (\ie an LRU miss that's a physical cache hit); once an LRU $i$ becomes less than $b_i$ it remains so until a future ripple eviction.

\section{Numerical results on cache\\ memory sharing}\label{sec:numer}

We ran a number of experiments to test our
working-set approximation of cache-hit
probabilities for the object-sharing cache.
To approximate,
we solved (\ref{t_i-equations}) using the
Newton-Raphson algorithm;
this was simplified by the concavity properties and uniqueness of
solution discussed in the proof of Prop.\ \ref{prop:approx}.

Typical  results for a  cache shared by
three or more LRU-lists are shown in
Tables \ref{table:shared3} and \ref{table:approx3}
(for $J=3$ LRU-lists).
Here we see that the approximation 
(\ref{L-def1})
is reasonably accurate. 

But for the $J=2$ LRU system, the approximation
(\ref{L-def1}) tends to underestimate cache-hit probabilities
by $\sim 30\%$.
One can use Jensen's inequality on (\ref{L-def1})  to get\footnote{By 
the same argument,
Proposition \ref{prop:approx} is also true
under (\ref{jensen}) or (\ref{L-def}),
the latter with $s_i>0$ for all $i$.}:
\be
L_{i,k}^{(1)} & \geq &  \ell_k \frac{1}{1+\sum_{j\not=i} h_{j,k}}=: L_{i,k}^* 
\label{jensen} \\
& \geq &  \ell_k \frac{h_{i,k}}{h_{i,k}+\sum_{j\not=i} h_{j,k}} =: 
L_{i,k}^{(2)}
\label{L-def}.
\ee
Empirically, we found that using $L_{i,k}^*$ for the working-set
approximation gives
approximate hitting probabilities only marginally larger than (\ref{L-def1}).
But, empirically, we found that using $L_{i,k}^{(2)}$ tends
to overestimate  
\cite{mcd-os}. 
That is, one can use (\ref{L-def1})
(\ref{L-def}) to find upper and lower bounds for the
$J=2$ cache case.



\begin{table}
\begin{tabular}{cccccccc} \hline
$i$ & $b_0$ & $b_1$ & $b_2$ & $h_{i,1}$ & $h_{i,10}$ & $h_{i,100}$ & $h_{i,1000}$\\ \hline
0& 8& 8& 8& 0.368& 0.0758& 0.0142& 0.00226\\ 
0& 8& 8& 64& 0.407& 0.0877& 0.0158& 0.00273\\ 
0& 8& 64& 8& 0.389& 0.0823& 0.0149& 0.00271\\ 
0& 8& 64& 64& 0.422& 0.0924& 0.0167& 0.0028\\ 
0& 64& 8& 8& 0.983& 0.5138& 0.1170& 0.02303\\ 
0& 64& 8& 64& 0.989& 0.5568& 0.1325& 0.02660\\ 
0& 64& 64& 8& 0.986& 0.5387& 0.1262& 0.02366\\ 
0& 64& 64& 64& 0.992& 0.5763& 0.1445& 0.02724\\ 
1& 8& 8& 8& 0.126& 0.0412& 0.0130& 0.00423\\ 
1& 8& 8& 64& 0.136& 0.0448& 0.0138& 0.00438\\ 
1& 8& 64& 8& 0.676& 0.2991& 0.1069& 0.03422\\ 
1& 8& 64& 64& 0.699& 0.3205& 0.1131& 0.03574\\ 
1& 64& 8& 8& 0.136& 0.0438& 0.0136& 0.00425\\ 
1& 64& 8& 64& 0.143& 0.0476& 0.0146& 0.00458\\ 
1& 64& 64& 8& 0.699& 0.3159& 0.1129& 0.03639\\ 
1& 64& 64& 64& 0.726& 0.3318& 0.1205& 0.03916\\ 
2& 8& 8& 8& 0.708& 0.1142& 0.0121& 0.00116\\ 
2& 8& 8& 64& 1.000& 0.7560& 0.1292& 0.01411\\ 
2& 8& 64& 8& 0.745& 0.1281& 0.0130& 0.00146\\ 
2& 8& 64& 64& 1.000& 0.7882& 0.1419& 0.01628\\ 
2& 64& 8& 8& 0.771& 0.1383& 0.0146& 0.00168\\ 
2& 64& 8& 64& 1.000& 0.7968& 0.1419& 0.01435\\ 
2& 64& 64& 8& 0.793& 0.1502& 0.0147& 0.00153\\ 
2& 64& 64& 64& 1.000& 0.8196& 0.1597& 0.01416
\end{tabular}
\caption{Empirical hitting probabilities for 
a simulated cache under the IRM of size $B=1000$ memory units
for unit-length objects  
($\forall n, ~\ell_n=1$) that is shared by three LRU-lists
$i=0,1,2$ respectively with Zipf popularity parameters
$\alpha_0=.75$, $\alpha_1=.5$, and $\alpha_2=1$. 
Simulation time was sufficiently long so
that these hitting probabilities are obtained with high 
confidence.
}\label{table:shared3}
\end{table}

\begin{table}
\begin{tabular}{cccccccc} \hline
$i$ & $b_0$ & $b_1$ & $b_2$ & $h_{i,1}$ & $h_{i,10}$ & $h_{i,100}$ & $h_{i,1000}$\\ \hline
0& 8& 8& 8& 0.365& 0.0776& 0.0143& 0.00255 \\ 
0& 8& 8& 64& 0.401& 0.0872& 0.0161& 0.00288 \\ 
0& 8& 64& 8& 0.386& 0.0832& 0.0153& 0.00274 \\ 
0& 8& 64& 64& 0.421& 0.0926& 0.0171& 0.00307 \\ 
0& 64& 8& 8& 0.984& 0.5213& 0.1228& 0.02302 \\ 
0& 64& 8& 64& 0.990& 0.5622& 0.1366& 0.02579 \\ 
0& 64& 64& 8& 0.988& 0.5455& 0.1308& 0.02463 \\ 
0& 64& 64& 64& 0.993& 0.5846& 0.1446& 0.02740 \\ 
1& 8& 8& 8& 0.126& 0.0416& 0.0133& 0.00424 \\ 
1& 8& 8& 64& 0.134& 0.0446& 0.0143& 0.00455 \\ 
1& 8& 64& 8& 0.678& 0.3011& 0.1071& 0.03519 \\ 
1& 8& 64& 64& 0.704& 0.3197& 0.1147& 0.03779 \\ 
1& 64& 8& 8& 0.133& 0.0442& 0.0142& 0.00451 \\ 
1& 64& 8& 64& 0.142& 0.0472& 0.0152& 0.00482 \\ 
1& 64& 64& 8& 0.701& 0.3171& 0.1136& 0.03742 \\ 
1& 64& 64& 64& 0.725& 0.3353& 0.1212& 0.04002 \\ 
2& 8& 8& 8& 0.694& 0.1116& 0.0118& 0.00118 \\ 
2& 8& 8& 64& 1.000& 0.7556& 0.1314& 0.01399 \\ 
2& 8& 64& 8& 0.734& 0.1242& 0.0132& 0.00133 \\ 
2& 8& 64& 64& 1.000& 0.7861& 0.1429& 0.01530 \\ 
2& 64& 8& 8& 0.756& 0.1314& 0.0140& 0.00141 \\ 
2& 64& 8& 64& 1.000& 0.7995& 0.1484& 0.01594 \\ 
2& 64& 64& 8& 0.787& 0.1434& 0.0154& 0.00155 \\ 
2& 64& 64& 64& 1.000& 0.8249& 0.1599& 0.01727  
\end{tabular}
\caption{Hitting probabilities numerically approximated instead using
mean object lengths (\ref{L-def1}), solving
(\ref{t_i-equations}) and substituting into (\ref{h_i-approx}),
for the shared cache of Table \ref{table:shared3}.
}\label{table:approx3}
\end{table}

Finally, we  note from Table \ref{table:approx4} the lower
cache-hit probabilities for a typical instance
of the set of parameters of the caching system of
Table \ref{table:shared3}, again consistent with the statement of
Prop.  \ref{prop:better-than-not-sharing}.
The differences range from marginal to over 10\% in the case of
the LRU $2$ with the smaller memory allocation ($b_2=8$).

\begin{table}
\begin{tabular}{cccccccc} \hline
$i$ & $b_0$ & $b_1$ & $b_2$ & $h_{i,1}$ & $h_{i,10}$ & $h_{i,100}$ & $h_{i,1000}$\\ \hline
0 & 64 & 64 & 8& 0.9800& 0.5084& 0.11760& 0.02259\\
1 & 64 & 64 & 8& 0.6683& 0.2944& 0.10437& 0.03503\\
2 & 64 & 64 & 8& 0.7005& 0.1123& 0.01176& 0.00113
\end{tabular}
\caption{Hitting probabilities of LRUs $i$
when caches are \underline{not} shared for
parameters of system of Table \ref{table:shared3}.
}\label{table:approx4}
\end{table}

\section{MemCacheD with Object Sharing (MCD-OS)}\label{sec:mcd-os}

\subsection{Background on Memcached}

Memcached (MCD) is 
a popular distributed in-memory cache~\cite{mcd-url} that offers a {\tt set/get} key-value
API (it offers some additional functions such as {\tt update} which is a special case of
{\tt set} so we ignore them).
Placement/routing of requests to servers within a cluster
is done via a consistent hashing function that clients apply to keys. 
If a {\tt get} request is a hit, the server holding the requested key-value
pair responds with the value. If the {\tt get} is a miss,
then the client must fetch the item from a (remote)
database and issue a {\tt set} command to
the cache. The {\tt set} command will add 
the object if it is not already in the cache,
otherwise it will update its value. To guarantee O(1) access time, 
MCD maintains a hash table on the server side linking all objects in cache, 
where an object is indexed by the hash value of its key. 

The basic unit of storage in MCD is an {\em item} which stores a key-value pair and some
meta-data such as a time-to-live (TTL) value. 
To overcome internal memory fragmentation, MCD divides memory into multiple {\em slabs} each of which
contains items within a range of sizes. Slabs are 1MB large by default. A group
of slabs containing items within the same size range is called a
{\em slabclass}. Instead of using the vanilla LRU, MCD uses type of 
segmented LRU (S-LRU) that is known to approximate LRU well while posing
lower computational needs (and processing delay) when servicing hits (which
is the common case in a well-provisioned cache). In MCD's S-LRU, items are
separated into three sub-lists called HOT, WARM, and COLD. Newly created
items always begin in HOT which is an LRU-based list. An item at the tail of 
HOT  is moved onto WARM only if it has a relatively long TTL and have been 
accessed at least twice (two or more accesses is taken as indicative of 
relatively high popularity). WARM holds popular and long-lived items and 
is operated as a first-in first-out (FIFO) list. Finally, COLD holds relatively
unpopular items and is operated as an LRU list.

\subsection{Our implementation}

We implement an MCD with object sharing, MCD-OS, by making modifications to Memcached v. 1.5.16. We make no changes to the client side of MCD. 
Our prototype is available here \cite{mcd-os}.
In particular, we retain MCD's consistent hashing functionality for 
client-driven content placement/routing in clustered settings as is. 
We make several changes to the server side of MCD. 
Requests coming from each proxy are handled by a pool of MCD-OS threads 
dedicated to that proxy. We retain the slabclass functionality for its 
fragmentation-related benefits and hash table for quick object access. An item's slabclass continues to be
determined by its actual (and not inflated/deflated) size. However, we remove 
per-slabclass LRU lists and instead implement a single LRU per proxy. 
Given our specific interest in the LRU replacement policy, we set up MCD-OS in 
our evaluation such that: (i) flat LRU as opposed to S-LRU is used and 
(ii) there is only one slabclass. Implementing MCD-OS for S-LRU with 
multiple slabclasses is part of our ongoing work. 

Note that on an LRU miss, MCD-OS will require the proxy to fetch the object
from a remote database and issue a {\tt set} command
to store it in cache followed by adding the item to the front of this 
proxy's LRU-list. Therefore,  there is no need for
MCD-OS to add an artificial delay in response to an
LRU miss that is a physical cache hit.

\begin{table}
\begin{tabular}{p{212pt}}
\toprule  
\textbf{proxy i issues {\tt get(k)}; hits in LRU i}\\
\midrule 
    \textbullet ~ promote item with key k to the head of LRU i \\
\midrule
\textbf{proxy i issues {\tt get(k)}; misses in LRU i but hits in cache}\\
\midrule
    \textbullet ~ insert the item with key k into the head of LRU i \\
    \textbullet ~ update the status of all other LRUs sharing this item 
    (deflation) \\
\midrule
\textbf{proxy i issues {\tt get(k)}; misses in both LRU i and cache }\\
\midrule
    \textbullet ~ return cache miss to client \\
                ~ // client is expected to fetch the item from database and issue {\tt set(k, v)}\\
\midrule
\textbf{proxy i issues {\tt set(k, v)}; key k doesn't exist in cache}\\
\midrule
    \textbullet ~ package the key-value pair {\tt (k,v)} into an item, store in cache \\
    \textbullet ~ set virtual length of the item to its actual length \\
    \textbullet ~ insert the item to head of LRU i \\
\midrule
\textbf{proxy i issues {\tt set(k, v)}; key k already exists in cache}\\
\midrule
    \textbullet ~ update the item with key k to reflect the new value {\tt v} \\
    \textbullet ~ promote the item to head of LRU i \\
    \textbullet ~ update the status of all other LRUs sharing this item (may involve a combination of inflation and deflation) \\
\bottomrule
\end{tabular}
\caption{Summary of MCD-OS behavior in response to {\tt set/get} requests from a proxy. }
\label{table:key_functionalities}
\end{table}

In Table~\ref{table:key_functionalities}, we summarize different types of behavior offered by MCD-OS in response to
{\tt set/get} requests from a proxy. We present the key functionalities implemented in MCD-OS to achieve
this behavior as a list of functions below.  We selectively list new logic added
by us and omit related functionality that MCD already implements. 
In the Appendix,
we provide detailed pseudocode for these
functions.  

\textbf{{\tt inflate}:} This new function is invoked when a shared item
 needs to be inflated. This happens upon the eviction of that item 
 from one of the proxy LRUs or if the virtual length of the item increases after a {\tt set} operation.  

  \textbf{{\tt deflate}:} This new function is invoked when a shared item  needs to be deflated. This happens upon the insertion into a proxy LRU of an item  that is shared with one or more other proxies, or if the virtual length of the item decreases after a {\tt set} operation. 

  \textbf{{\tt insert}:} This is analogous to the native MCD function 
{\tt item\_link} 
that inserts an item into the appropriate LRU-list. It is used for
  item insertion and replacement. We modify it to also invoke the functions {\tt inflate} or {\tt deflate} corresponding to
  an increase or a decrease in the virtual length of the inserted item.

  \textbf{{\tt evict}:} This is analogous to the native MCD function 
{\tt item\_unlink} that evicts an item from its LRU-list. We modify it to also
  invoke the function {\tt inflate} after item eviction to update virtual lengths of copies of the evicted item that
  still resides in some other proxies' LRU-lists. 

  \textbf{{\tt process\_command}:} This is a native MCD function that parses client requests and implements {\tt get} and {\tt set}
  logic. We enhance it to additionally implement object sharing. 


\subsection{Overhead of object sharing for MCD-OS}\label{sec:ripple-overhead}

Object sharing introduces additional overhead for {\tt set} commands
associated with  a ripple of evictions among the LRUs
owing to item size deflation/inflation.
In the following, we compare the overhead of
{\tt set} commands (after a cache miss)
for MCD-OS and MCD, the latter with the
same collective {\tt get} commands but a single LRU cache of
the same collective size ($\sum_i b_i$).

\begin{figure}[!t]
    \includegraphics[width=\linewidth]{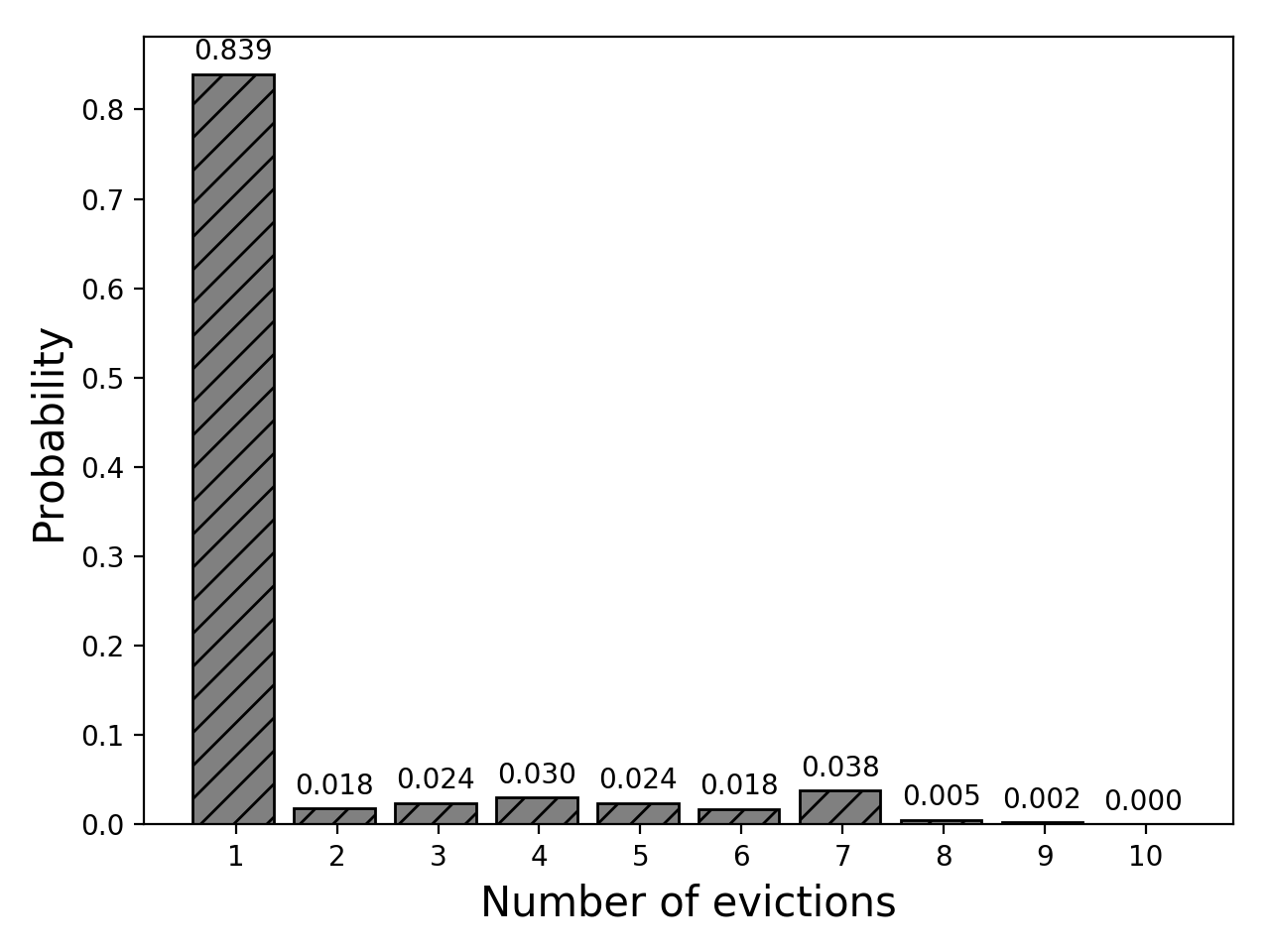}
    \caption{A histogram of the number of evictions per {\tt set} request under MCD-OS. There were no {\tt set} commands observed with more than 10 associated evictions. Note that the number for MCD without object sharing is always 1.}\label{fig:ripple}
\end{figure}

For our experiments, we used $J=9$ proxies with $N=10^6$ items,
where each item was 100kB.  
The total cache memory was 3 GB. 
In a typical experiment, we considered 
very different proxies $i\in[J]$ with
Zipf parameter $0.5+0.5(i-1)$ and  memory allocations:
$b=$100 MB for proxies 1,2,3;
$b=$200 MB for proxies 4,5,6; and
$b=$700 MB for proxies 7,8,9.
The number of {\tt get} commands issued in each experiment was
$3\times 10^6$ after the cold misses have abated.
The 
histogram 
of the number of evictions per {\tt set} request
for MCD-OS is given in Figure \ref{fig:ripple}. As shown, in a small
number of cases, the size of this ``eviction ripple'' can be as large
as 9. However, overall only 16\% of the {\tt set} requests experienced
more than one eviction (i.e., an overhead beyond what an eviction in MCD
would experience).

\begin{figure}[!t]
    \includegraphics[width=\linewidth]{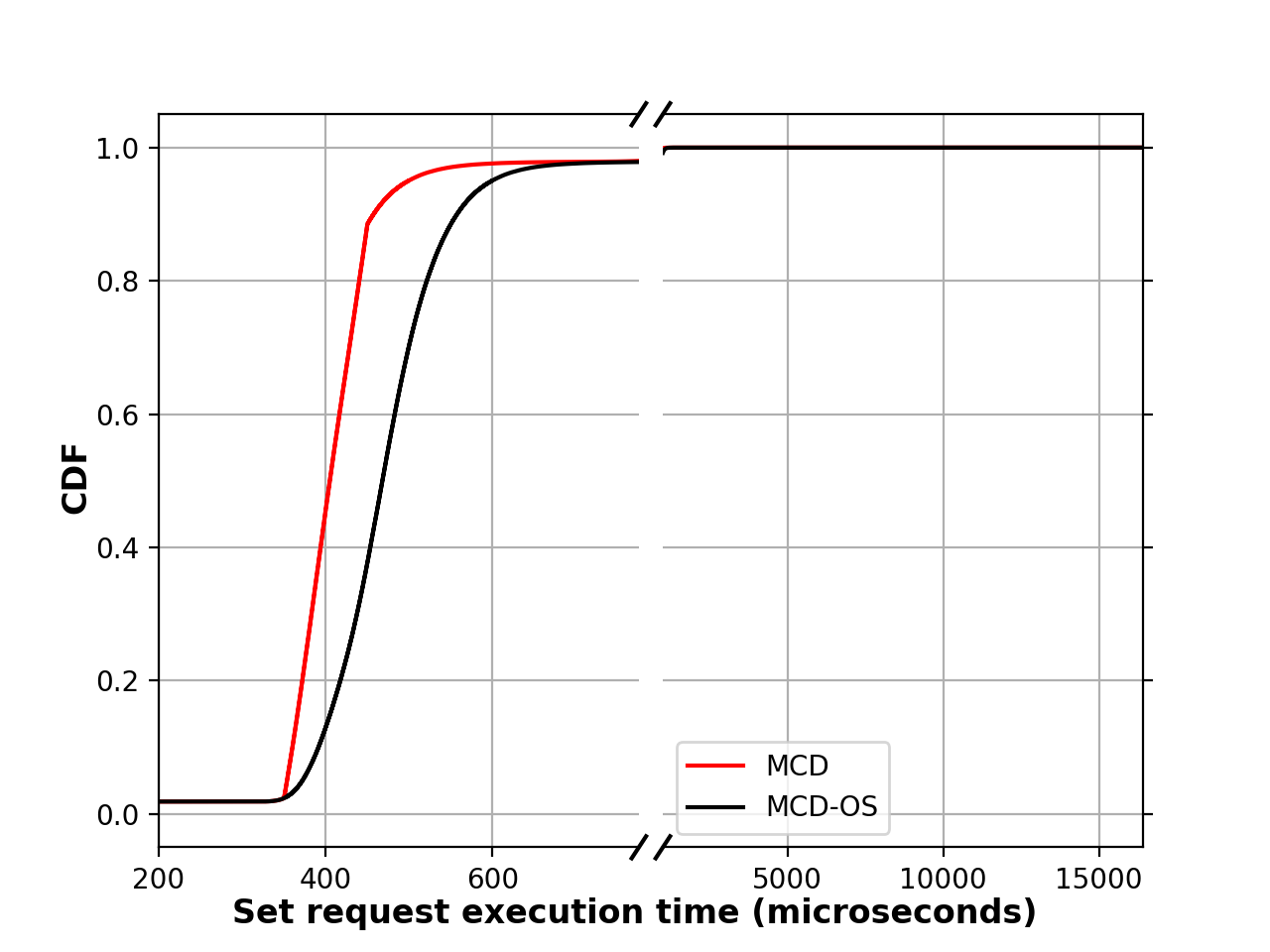}
    \caption{CDFs of the {\tt set} request execution times
comparing MCD with MCD-OS.}\label{fig:cdf}
\end{figure}

In Figure \ref{fig:cdf}, the CDFs of the {\tt set} execution times
are plotted under both MCD and MCD-OS. Note that, though there
is a single eviction per {\tt set} under MCD, there is some variability
when updating the LRU. 
See Table \ref{tab:set_stats}.

\begin{table}
\begin{center}
\begin{tabular}{c|c|c}
\hline
cache & mean & std dev\\ \hline
MCD & 412 $\mu$s & 111 $\mu$s\\ \hline
MCD-OS & 474 $\mu$s & 127 $\mu$ s  \\ \hline
\end{tabular}
\end{center}
\caption{Means and standard deviations of {\tt set} request
execution times under MCD-OS and MCD.}\label{tab:set_stats}
\end{table}


Other experiments showed that when all the proxies are very similar,
 the additional {\tt set} overhead was reduced, even negligible.
Also, a {\tt get} under MCD-OS would obviously require additional 
overhead to look-up which LRU (based on proxy identifier)
is requested, but  we found it to be negligible.

\section{Summary and Future Work}\label{sec:concl}

In this paper, we considered object sharing by LRU caches. Such sharing
will reduce the cost of operation at a given level of performance
(cache-hit probabilities) or improve performance for given budgets.
We proposed an extension of the classical working-set approximation
of cache-hit probabilities to this shared-object setting, and evaluated
its performance both numerically and 
based on experiments with a Memcached prototype
(MemDacheD-OS or MCD-OS) \cite{mcd-os}. 
This approximation may be used for admission control to 
help determine virtualized memory resources to be allocated 
under object sharing.
We also
numerically evaluated the {\tt set}
overhead of object sharing.

We have also implemented MCD-OS for commonly used Segmented-LRU (S-LRU)
with multiple slabclasses, where S-LRU is designed
to reduce  memory overhead for popular (hot) objects.  
Cache-hit probabilities do  not change significantly 
($\sim 2-3\%$ difference) under S-LRU
under object sharing.

In ongoing work, we are evaluating the method of 
Section \ref{sec:os-rre} to reduce the overhead of ripple evictions.
We are also evaluating MCD-OS with variable-length objects
which are allocated in different slabs.

\bibliographystyle{plain}
\bibliography{../latex/cache,../latex/games,../latex/quasi-rev,../latex/sim_ann,../latex/wiley,../latex/scheduling,../latex/stochastic,../latex/positively-associated}

\begin{thebibliography}{10}

\bibitem{basar99dynamic}
T.~Ba\c{s}ar and G.J. Olsder.
\newblock {\em Dynamic Noncooperative Game Theory}.
\newblock Classics in Applied Mathematics, SIAM, Philadelphia, 1999.

\bibitem{Border85}
K.C. Border.
\newblock {\em Fixed Point Theorems with Applications to Economics and Game
  Theory}.
\newblock Cambridge University Press, London, 1985.

\bibitem{Che02}
H.~Che, Y.~Tung, and Z.~Wang.
\newblock {Hierarchical Web Caching Systems: Modeling, Design and Experimental
  Results}.
\newblock {\em IEEE JSAC}, 20(7), Sept. 2002.

\bibitem{Towsley17}
M.~Dehghan, W.~Chu, P.~Nain, and D.~Towsley.
\newblock {Sharing LRU Cache Resources among Content Providers: A Utility-Based
  Approach }.
\newblock {\em IEEE/ACM Transactions on Networking (TON)}, 27(2), Apr. 2019.

\bibitem{DS72}
P.J. Denning and S.C. Schwartz.
\newblock Properties of the working-set model.
\newblock {\em Commun. ACM}, 15(3):191--198, March 1972.

\bibitem{Eyilmaz19}
A.~Eryilmaz and al.
\newblock {A New Flexible Multi-flow LRU Cache ManagementParadigm for
  Minimizing Misses}.
\newblock In {\em Proc. ACM SIGMETRICS}, 2019.

\bibitem{Fagin77}
R.~Fagin.
\newblock Asymptotic approximation of the move-to-front search cost
  distribution and least-recently-used caching fault probabilities, 1977.

\bibitem{Fricker12}
C.~Fricker, P.~Robert, and J.~Roberts.
\newblock {A Versatile and Accurate Approximation for LRU Cache Performance}.
\newblock In {\em Proc. International Teletraffic Congress}, 2012.

\bibitem{Golrezaei12}
N.~Golrezaei, K.~Shanmugam, A.G. Dimakis, A.F. Molisch, and G.~Caire.
\newblock {Femtocaching: Wireless video content delivery through distributed
  caching helpers}.
\newblock In {\em Proc. IEEE INFOCOM}, 2012.

\bibitem{J-DP83}
K.~Joag-Dev and F.~Proschan.
\newblock Negative association of random variables with applications.
\newblock {\em The Annals of Statistics}, pages 286--295, 1983.

\bibitem{mcd-os}
G.~Kesidis, N.~Alfares, X.~Li, B.~Urgaonkar, and M.~Kandemir.
\newblock {Working-Set Approximation for a Caching System with Object Sharing}.
\newblock https://github.com/PSU-Cloud/MCD-OS/, Aug. 2019.

\bibitem{KS81}
A.~Khursheed and K.M.L. Saxena.
\newblock Positive dependence in multivariate distributions.
\newblock {\em Communications in Statistics - Theory and Methods},
  10(12):1183--1196, 1981.

\bibitem{mcd-url}
Memcached.
\newblock https://memcached.org/.

\bibitem{Moulin84}
H.~Moulin.
\newblock {Dominance Solvability and Cournot Stability}.
\newblock {\em Mathematical Social Sciences}, 7:83--102, 1984.

\bibitem{Tassiulas19}
K.~Poularakis, G.~Iosifidis, A.~Argyriou, I.~Koutsopoulos, and L.~Tassiulas.
\newblock {Distributed Caching Algorithms in the Realm of Layered Video
  Streaming}.
\newblock {\em IEEE Trans. Mob. Comput.}, 18(4):757--770, 2019.

\bibitem{Stoica16}
Q.~Pu, H.~Li, M.~Zaharia, A.~Ghodsi, and I.~Stoica.
\newblock {FairRide: Near-Optimal, Fair Cache Sharing}.
\newblock In {\em Proc. USENIX NDSI}, Santa Clara, CA, USA, March 2016.

\bibitem{Rosen65}
J.B. Rosen.
\newblock Existence and uniqueness of equilibrium points for concave $n$-person
  games.
\newblock {\em Econometrica}, 33(3):520--534, 1965.

\bibitem{squid}
{Squid: Optimising Web Delivery}.
\newblock http://www.squid-cache.org.

\bibitem{Wajc17}
D.~Wajc.
\newblock Negative association: Definition, properties and applications.
\newblock http://www.cs.cmu.edu/$\sim$dwajc/notes/Negative\%20Association.pdf,
  Apr. 2017.

\bibitem{YWang16}
Y.~Wang, X.~Zhou, M.~Sun, L.~Zhang, and X.~Wu.
\newblock {A new QoE-driven video cache management scheme with wireless cloud
  computing in cellular networks}.
\newblock {\em Mobile Networks and Applications}, 2016.

\end{thebibliography}


\end{document}